\documentclass[%
 reprint,
 amsmath,amssymb,
 aps,
]{revtex4-2}

\usepackage{graphicx}% Include figure files
\usepackage{dcolumn}% Align table columns on decimal point
\usepackage{bm}% bold math
\usepackage{booktabs}
\usepackage{xspace}
\usepackage{textcomp}
\usepackage{xcolor}
\usepackage{tabularx}

\begin{document}

\preprint{APS/123-QED}

\title{Influence of Conical Wire Array Geometry on Flow and Temperature Profiles Measured via Thomson Scattering and Optical Techniques}
%Flow and Temperature Profiles of Conical Wire Array Jets Measured via Thomson Scattering and Optical Techniques}
%Investigating pulsed power driven plasma jets using Thomson scattering, Stark broadening and laser probing}

\author{Luisa Izquierdo$^{1,*}$, Felipe Veloso$^{1,\dagger}$, Miguel Escalona$^1$, Vicente Valenzuela-Villaseca$^2$, Gonzalo Avaria$^3$, Julio Valenzuela$^1$}

\affiliation{$^1$Instituto de Física, Pontificia Universidad Católica de Chile, Santiago 7820436, Chile. \\
             $^2$Department of Astrophysical Sciences, Princeton University, Princeton, New Jersey 08544, USA.\\
             $^3$Departamento de Física, Universidad Técnica Federico Santa María, Santiago 8940897, Chile}

\email{luisa.izquierdo@uc.cl,$\dagger$fveloso@uc.cl}

%Parametric characterization of pulsed-power-driven plasma jets}% 

%\author{}%
% \email{}
%\affiliation{%
% Pontificia Universidad Católica de Chile, Santiago, Chile.


%\affiliation{
%}


\begin{abstract}

%This work presents a detailed experimental 
Conical wire arrays with different opening angles are used as load of a 400kA, 1kA/ns generator. The differences in opening angle allow the study of the influence of the array geometry on the jet properties. The characterization of the jets is performed using a combination of advanced diagnostic techniques, including moiré schlieren deflectometry, visible self-emission spectroscopy, and optical Thomson scattering. The results reveal that, under the experimental conditions, the plasma jets exhibit electron temperatures ranging from $8$ to $17$ eV, increasing along the axial direction. In contrast, the ion temperature decreases from approximately $35$ eV near the base of the jet to about $20$ eV at higher axial positions. The electron density profile peaks at $\sim 4 \times 10^{18}$ cm$^{-3}$ in the central lower region of the jet and decreases with height exponentially with a characteristic lenght $L_n = $2.86 mm. This behavior is reproducible and independent of the conical array geometry. However, the cone opening angle significantly affect the jet propagation velocity, with larger opening angles producing higher axial velocities ($V_{\phi=40^\circ} \approx 125\pm3$ km/s, $V_{\phi=20^\circ} \approx 98\pm5$ km/s), demonstrating that the cone geometry provides effective control over the jet propagation velocity.
\end{abstract}

\maketitle

%\tableofcontents

\section{Introduction}

Astrophysical plasma jets \cite{de1991astrophysical} are ubiquitous phenomena observed across a wide range of environments, including active galactic nuclei (AGNs) \cite{burrows1996hubble}, young stellar objects (YSOs) \cite{reipurth2001herbig}, and planetary nebulae (PNs) \cite{weiler1988supernovae}. These jets typically manifest as highly collimated, supersonic plasma outflows, playing critical roles in the transport of mass, energy, and magnetic fields over large astrophysical scales.

Motivated by the astrophysical relevance of plasma jets, a variety of laboratory platforms have been developed to generate and study scaled analogs of these phenomena. Experiments based on conical wire arrays \cite{izquierdo2022shock, lebedev2002laboratory, lebedev2005production}, radial wire arrays \cite{ciardi2009episodic, bott2015investigation}, and gas-puff Z-pinches \cite{lavine2025experimental} have all successfully produced radiatively cooled, magnetically collimated plasma jets that satisfy scaling laws relevant to astrophysical systems. These laboratory experiments have enabled the study of jet propagation, stability, episodic ejection, interaction with ambient media, and radiative cooling under controlled conditions.

Despite differences in experimental implementation, many of these platforms share common features: the generation of a dense plasma column, the presence of a surrounding magnetic field that assists in jet collimation, and the occurrence of radiative cooling that influences jet morphology. These similarities have allowed experimental studies to explore universal aspects of plasma jet physics across a range of conditions.

Among these platforms, plasma jets produced by conical wire array Z-pinches stand out for their high degree of collimation and relevance to young stellar object outflows \cite{lebedev2005production, veloso2016plasma}. However, despite significant advances, experimental characterizations of these laboratory plasma jets have largely relied on indirect diagnostic methods. Most notably, time-resolved laser probing (e.g., interferometry and schlieren) has been used to infer axial density variations, while self-emission imaging has provided qualitative information about jet morphology and dynamics \cite{lebedev2002experiments,veloso2016plasma}. These techniques, although powerful, do not directly measure key plasma parameters such as the flow velocity or internal gradients, often leading to ambiguities in the interpretation of jet structure. For instance, axial variations observed in interferometry can result from either real flow dynamics or geometric zippering effects, and numerical simulations predict axial velocity gradients \cite{ciardi2002modeling} that remain unobserved experimentally.

In addition to diagnostic limitations, the standard conical wire array configuration allows significant lateral plasma inflows toward the axis during jet formation. These inflows, which can contribute to the closure of electrical current through the plasma column \cite{ciardi2007episodic, suzuki2013interaction}, complicate the interpretation of experimental results and may artificially enhance jet stability and density profiles.

In this work, to address these questions, we performed a detailed experimental study of plasma jets emitted by aluminum conical wire arrays driven by the Llampudken pulsed power generator. This study combines Thomson scattering, optical self-emission spectroscopy, and moiré schlieren deflectometry to spatially resolve plasma density, temperature, and flow velocity. Moreover, to improve jet purity and reproducibility, an aperture was incorporated into the upper electrode to filter out parasitic plasma flows originating from the wire ablation phase before the formation of the main jet. Finally, by varying the opening angle of the conical array, we investigate how geometrical factors influence the plasma parameters of the emitted jets, providing new insights into the formation mechanisms and flow dynamics relevant to laboratory astrophysics experiments.

\section{Experimental Setup and Diagnostics}

\begin{figure}[h]
    \centering
    \includegraphics[width=0.5\textwidth]{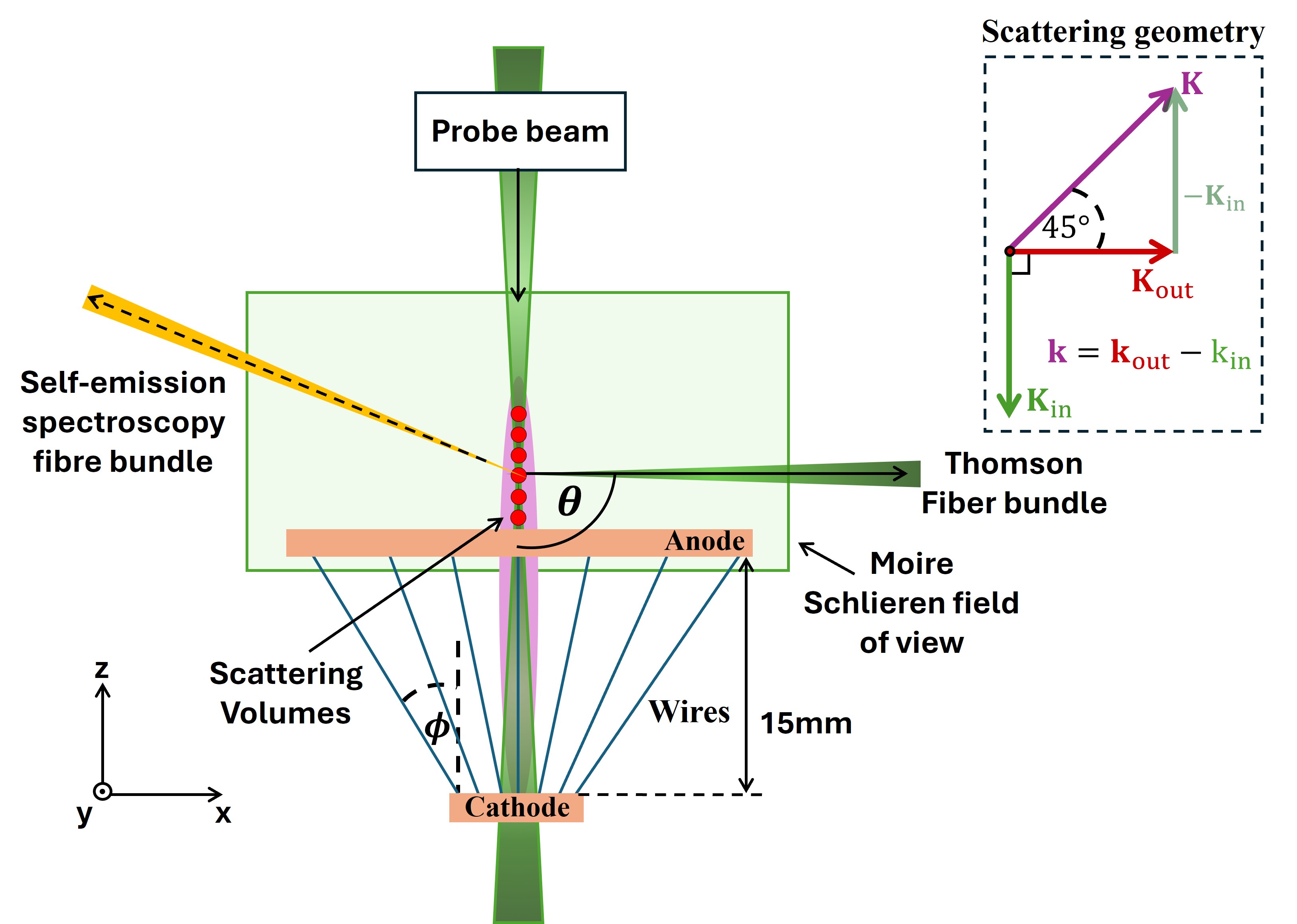}
    \caption{Schematic diagram of experimental setup. Conical wire array opening angle is determined by the angle $\phi$. TS volumes are overlaid in red circles (not scale). Continuous arrows indicate the scattering geometry. Dashed arrows show the collection direction of the spectroscopy. Additionally, the field of view of side-on moiré schlieren.}
    \label{setup}
\end{figure}

\noindent Plasma jets are generated when the plasma ablated from the wires is accumulated at the axis of the configuration, and later expelled axially due the zippering effect of the conical wire array when used as the load of the $\sim 400$ kA,
$\sim 1$kA/ns current pulse of the Llampudken pulsed-power generator \cite{Llampudken}. The conical wire array consists of sixteen equally spaced aluminium wires of $50$ $\mu$m diameter each. The total height of the array is given by interelectrode separation, which is set constant at 15mm in all the experiments reported here. A schematic of the experimental setup is shown in Fig. \ref{setup}.

To experimentally investigate the effect of the opening angle of the conical wire array ($\phi$, measured with respect to the vertical, as shown in Fig. \ref{setup}), three different angles are studied: $\phi = 20^\circ$, $30^\circ$, and $40^\circ$. 
In all these cases, the cathode diameter remains fixed at $7$ mm while the anode diameter is adjusted accordingly to achieve the desired opening angles. Hence,  anode diameters of $17.9$ mm, $24.3$ mm, and $32.1$ mm, are used corresponding to the respective opening angles. 
In addition, the array load is covered by a $3$ mm thick metallic lid having a central aperture of 5mm in diameter. This aperture is geometrically designed for limiting the plasma emerging above the interelectrode region (above anode). Using this aperture, the ablated plasma emerging from the wires driven by the Lorentz force (i.e., plasma flares propagating perpendicular to the wires) cannot pass directly through the lid aperture. This is true for the $\phi = 20^\circ$ and $\phi = 30^\circ$ cases. For the $\phi = 40^\circ$, the ablated plasma can reach at most $2.09$ mm over the anode lid. On the contrary, the plasma jet can pass through the central aperture due to its axial propagation. Therefore, the measurements performed above anode (lid) surface correspond exclusively to the plasma jet and not to the ablation flares emitted from the wires.
Throughout this article, temporal reference $t=0$ is considered when current begins to flow through the load. Moreover, axial propagation of the jet is measured by considering $z=0$ as the anode surface above the interelectrode region.

In the experiments, the overall structure and plasma parameters are investigated using a comprehensive multi-diagnostic suite. Moiré schlieren deflectrometry \cite{hutchinson} is used to measure the line integrated electron density gradient in the side on direction. It is carried out using the second ($532$ nm) harmonic of a Nd:YAG laser. The moiré fringe pattern is formed with two Ronchi gratings with a ruling of 20 lines/mm, separated by $1.57$ cm, corresponding to three Talbot distances and rotated by an angle $\psi=10^\circ$ with respect to each other \cite{julio-moire}. Structures in the moiré schlieren images %corresponding to the period of the gratings
are eliminated using spatial Fourier filters that selects the frecuencies corresponding to the moiré pattern. The shift of moiré fringes is proportional to the line integrated electron density gradient $\int \nabla n_e dl$ and the volumetric electron density $n_e$ can be reconstructed using an onion-peeling inversion method \cite{dasch1992one}, considering axial symmetry. In this setup, electron densities above $6.7\times10^{17}$ cm$^{-3}$ can be measured. As shown in Fig.\ref{setup}, the laser beam is aligned side-on to the experiment, passing over the top edge of the upper electrode to obtain a cross-sectional view of the plasma jet.

\begin{figure*}[ht]
    \centering
    \includegraphics[width=\textwidth]{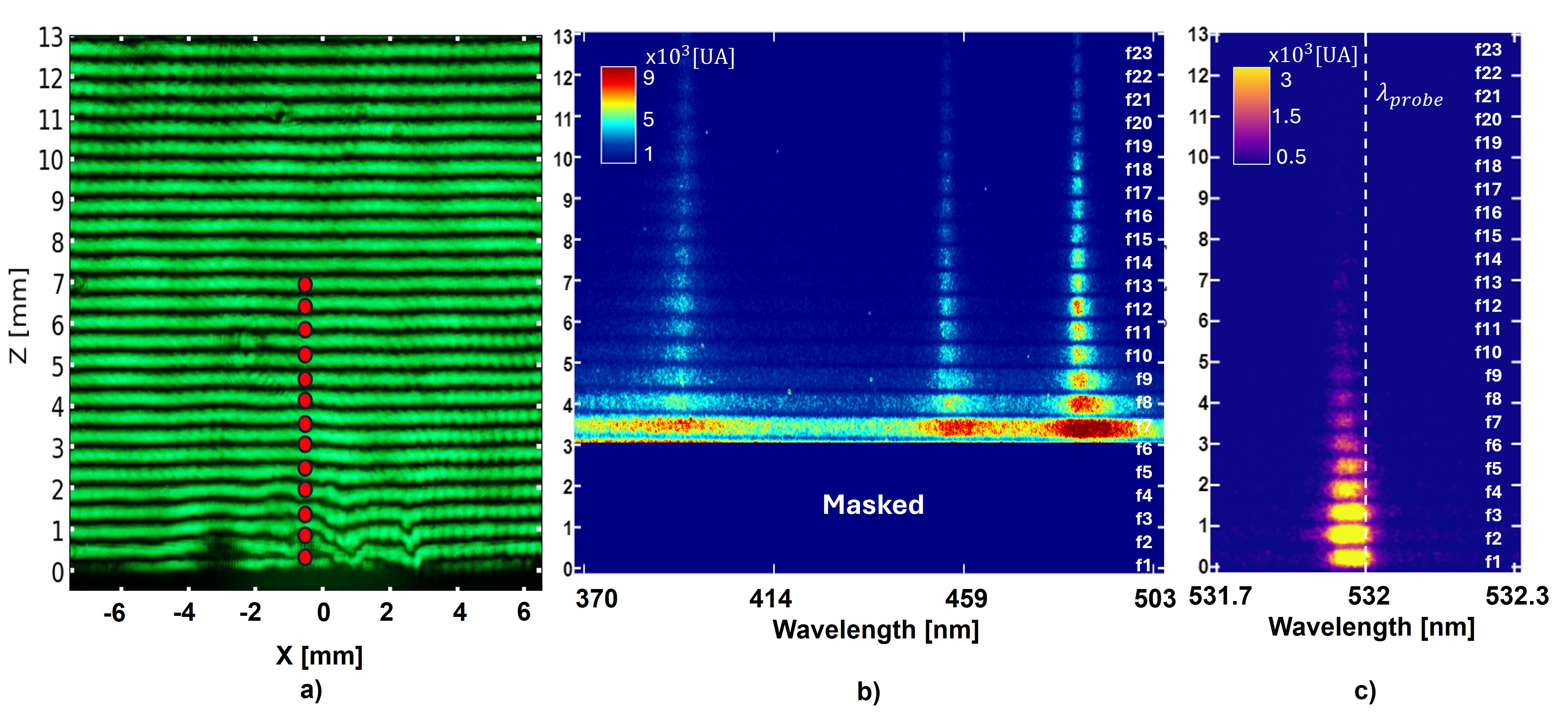}
    \caption{Raw experimental data of a plasma jet emitted by a $20^o$ conical wire array at $450$ns. TS volumes are overlaid in red circles into the side on moiré pattern a). The information was measured just above the anode of the conical wire array ($Z=0$). b) Visible self-emission spectrum. c) Thomson scattering spectrum of the fiber bundle A. white dashed line indicate the probe wavelength.}
    \label{raw data}
\end{figure*}

In addition, %to moiré schlieren deflectometry,%
visible self-emission spectroscopy and optical Thomson Scattering (TS) are employed, both providing spatial resolution along the axial propagation direction of the plasma jet. TS is used to measure the flow velocity and electron and ion temperatures. A laser beam ($532$ nm, $1$ J, $4$ ns) is focused along the plasma jet propagation axis and scattered light was colected by a fiber optic bundle with an scattering angle $\theta = 90^\circ$, as shown in Fig.~\ref{setup}. The fiber bundle consist of a linear array of $25$ fiber optics ($200$ $\mu$m diameter, $200$ $\mu$m fiber to fiber distance), and light is collected at $25$ distinct axial positions. The Thomson scattering collection volume is determined by the cross-sectional area of the laser beam and the optical collection volume length, given by $l = 2.1 \times 200$ $\mu$m $= 420$ $\mu$m. The collected light is transmitted to a $500$ mm focal length spectrometer (SpectraPro HRS-500) equipped with a $50$ $\mu$m entrance slit and a $2400$ l/mm grating. Spectra are recorded using a gated ICCD camera (Stanford 4 Picos) with an exposure time of $6$ ns.

%the scattered light was collected from 25 equally spaced positions using a lens and a linear array of optical ﬁbers, with 420um diameter collection volumes spaced by 210 um.

%Brewster windows were installed at the chamber's input and output to minimize stray light contamination in the TS spectra.

%implemented using a Nd:YAG laser (EKSPLA NL310) at $\lambda=532$ nm, capable of delivering pulses of up to $1$ J with a full width at half maximum (FWHM) of $4$ ns. As shown in Fig.~\ref{setup}, the laser beam was focused along the plasma jet propagation axis. Brewster windows were installed at the chamber's input and output to minimize stray light contamination in the TS spectra.

%Scattered light was collected at $\theta = 90^\circ$ with respect to the laser beam (see Fig.~\ref{setup}), using a multimode linear fiber array with a magnification of $\sim 2.1$. The fiber bundle consisted of $25 \times 200$ $\mu$m diameter fibers, aligned along the jet propagation path. The Thomson scattering collection volume was determined by the cross-sectional area of the laser beam and the optical collection volume length, given by $l = 2.1 \times 200$ $\mu$m $= 420$ $\mu$m. The collected light was transmitted to a $500$ mm focal length spectrometer (SpectraPro HRS-500) equipped with a $50$ $\mu$m entrance slit and a $2400$ l/mm grating. Spectra were recorded using a gated ICCD camera (Stanford 4 Picos) with an $8$ ns time window. The resulting scattering vector diagram is shown in Fig.~\ref{setup}.

Similarly, self-emission spectra are acquired using a linear fiber bundle positioned to observe the same regions diagnosed with TS. The fiber bundle consists of $25 \times 200$ $\mu$m diameter fibers, aligned vertically above the anode of the conical wire array along the jet propagation path, with an approximate magnification of $2.1$. The collected emission is analyzed using a $500$ mm focal length spectrometer equipped with a $150$ l/mm grating. Spectra were recorded using a gated ICCD camera (Andor iStar) with an exposure
time of 3 ns, covering the wavelength range $365-510$ nm, where Al-III spectral lines are present.

\section{Results}
Fig. \ref{raw data} shows raw data of a plasma jet emitted by a $20^o$ conical wire array, at $t \sim 450$ ns after current discharge. 
By combining the results of different diagnostics, a plasma jet characterization up to $14$ mm above anode surface is achieved. 

\begin{figure}[]
    \centering
    \includegraphics[width=0.45\textwidth]{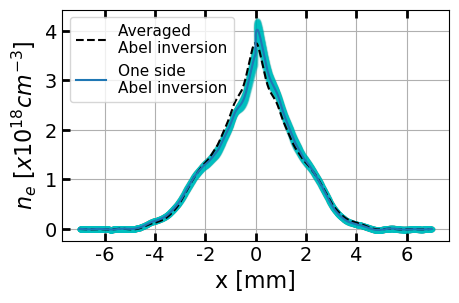}
    \caption{Abel inverted electron density obtained with moiré schlieren deflectometry measurement at $\sim 450$ ns plasma jet of a $20^o$ conical array.}
    \label{ps tnt}
\end{figure}

%with an extension of approximately $13$ mm is observed. 
Raw deflectometry data (Fig.\ref{raw data}a) is analyzed using the TNT code \cite{perez2022tia,TNT} to obtain a 1D phase shift of the moiré pattern. 
%Figure \ref{ps tnt}a presents the phase shift measured at $z\approx 1$ mm (fringe 3) of the moiré deflectometer. Differences in the phase shift profile are observed between the two sides of the jet's propagation axis ($x=0$). To assess the degree of cylindrical symmetry, Fig. \ref{ps tnt}b shows the corresponding volumetric electron density for each side, assuming symmetry, as well as the average of both sides. Deviations from axisymmetry are quantified by comparing individual inversions with this averaged profile. The difference between the individually inverted electron density profiles and the averaged profile has a characteristic deviation of approximately $5\%$.
In order to assess cylindrical symetry, we analize moiré schlieren at both sides of the symetry axis individually. As shown in Figure \ref{ps tnt}, the average electron density profile differ less than $5\%$ from each side, demonstrating that cylindrical symetry assumption is valid. Our moiré measurements reveal a radial electron density profile that peaks in the propagation axis in $n_e^{\phi=20^\circ} (t=450\text{ ns})=(4.0\pm0.2)\times10^{18}$ cm$^{-3}$ and radially decrease.

%%%%%%%%%%%%%%%%%%%
% ¿porque poner t\sim 450ns y no t=450ns?
% deberiamos poner =
%%%%%%%%%%%%%%%%%%%
%The TNT phase shift map of the integrated gradient electron density shown in Fig. \ref{ps tnt}a was Abel-inverted using \ref{abel moire} with an onion-peeling method applied independently to each side of the propagation axis. The resulting electron density profile is presented in Fig. \ref{ps tnt}b. This profile reveals a maximum electron density of approximately $\sim 4\times 10^{18}$ $cm^{-3}$ in the center of the jet. 
Nevertheless, density measurements using moiré deflectometry are limited by the ability to detect fringe shifts in the pattern. To extend the axial range of density determination, we complement moiré data with spectroscopy. 
Figure \ref{raw data}b presents raw self-emission spectral data of the plasma jet, where three emission lines corresponding to Al-III transitions can be clearly distinguished. It is observed that the width of the emission lines varies with the axial position. The lines are broader at lower axial positions (closer to the anode) and gradually narrow at higher positions along the $z$-axis. This behavior indicates a decreasing trend in electron density as a function of height.

\begin{figure}[h]
    \centering
    \includegraphics[width=0.45\textwidth]{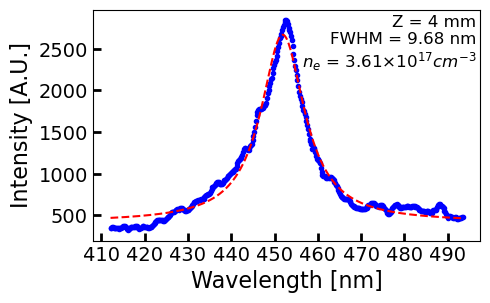}
    \caption{Characteristic spectral emision line at 452.8 nm (blue dots). Red dashed lines show the lorentzian fit used to determine the FWHM used to estimate stark broadening.}
    \label{stark FWHM}
\end{figure}

The full width at half maximum (FWHM) of the Al-III emission line at $452.9$ nm is measured using a Lorentzian fit to the spectral data (red dashed lines in Fig. \ref{stark FWHM}). After correction for instrumental broadening, local electron density values are obtained as a function of axial position by evaluating the Stark broadening relation in the regime where broadening is dominated by electron collisions \cite{Hahn, Huddlestone}. A Stark parameter of $\omega = 0.134 \pm 0.025$ pm was used \cite{dojic2020stark}.

\begin{figure}[]
    \centering
    \includegraphics[width=0.48\textwidth]{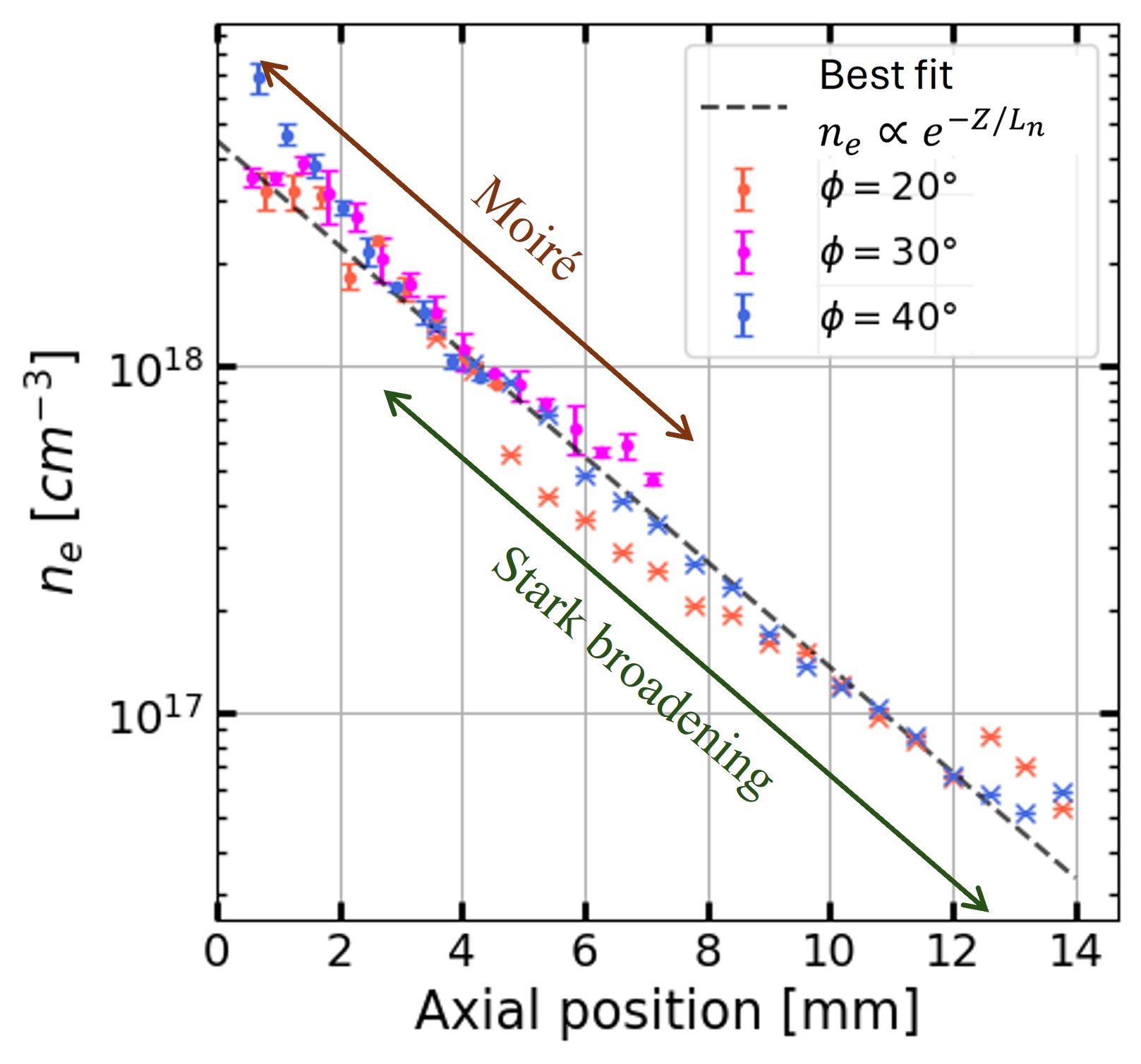}
    \caption{Peak electron density profile as a function of axial position ($z$) at the jet axis.
    %x-position of the fiber bundles 
    for three different conical wire array opening angles. Dot markers represent measurements obtained using moiré Schlieren deflectometry, while x-markers indicate results from Stark broadening. The dashed black line shows the best fit to the $20^\circ$- cone data, given by $n_e = 4.5e^{-0.35z}$.}
    
    \label{ne todos los angulos}
\end{figure}

By combining the deflectometry and spectroscopy method, the peak electron density as a funcion of the axial position (i.e., $n_e^{max}(z)$) is obtained and shown in Fig. \ref{ne todos los angulos}. Electron density values obtained from both methods exhibit good agreement in the overlapping measurement regions, which indicate consistency in the applied methods.
%thus validating each other.

From Fig \ref{ne todos los angulos}, it can be seen that peak electron density reaches maximum at the base of the jet near the anode (i.e., closer to the conical array load) 
%
%For all opening angles studied and in line with the qualitative observations from the deflectometry images and the self-emission spectrum, the axial electron density profile reveals that the electron density peaks at lower axial positions, 
reaching a maximum of approximately $4 \times 10^{18}$ cm$^{-3}$. At higher axial positions, the electron density decreases, reaching approximately $6 \times 10^{16}$ cm$^{-3}$ at $z \approx 13$ mm. 
Moreover, peak density decays exponentially with axial position as $n_e^{max}(z)\propto e^{-z/L_n}$ with $L_n=2.86$ mm. Therefore we consider $L_n$ as a caharacteristic xial density gradient scale length. Furthermore, electron density remains within the same magnitude regardless of the opening angle $\phi$.
%, which can be considered as the characteristic lenght of the jet.
%
%{\color{red} OJO!!! CON ESTE $z_0$ podemos justificar el porque usar z=3mm en parametros adimensionales!!!(de hecho, sería mejor usar (2.86 +/- algo) mm)}

%
%The axial electron density profile exhibits an exponential decay, following a relation $n_e \propto e^{-a\cdot z}$, where $a$ is a constant.
%This behavior remains consistent for all $\phi$. Furthermore, the electron density does not vary significantly with the opening angle, with a difference factor between the densities measured at the same position $< 2$.

%The ion acoustic feature of TS provided localized measurements of plasma velocity and temperatures at 10 localized plasma volumes (shown in Fig. \ref{raw data}a). 
On the other hand, TS spectra are analyzed using a bayesian inference method, as described in \cite{escalona2023bayesian}. Using these data, velocity and temperature measurements were obtained from 10 scattering volumes (red circles in the Fig. \ref{raw data}a).

\begin{figure}[]
    \centering
    \includegraphics[width=0.45\textwidth]{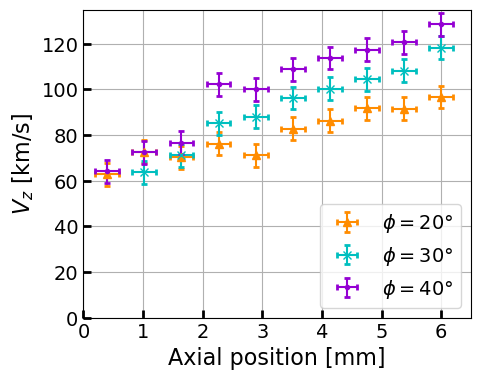}
    \caption{Axial velocity profile of the plasma jet for different opening angles, ~450ns after the beginning of current discharge}
    \label{velocity profiles}
\end{figure}

The flow velocity of the plasma jet was determined from the Doppler shift of the scattered spectra. As shown in Fig.\ref{velocity profiles}, the flow velocity increases approximately linearly with the axial position, a behavior observed consistently across all opening angles studied. The rate of increase in flow velocity with axial position is higher for larger $\phi$, with an increment of approximately $3$ km s$^{-1}$/mm for every $10^\circ$ increase in $\phi$. This results in a maximum velocity of $125\pm3$ km/s at $z = 6$ mm for $\phi = 40^\circ$, compared to $98\pm5$ km/s at the same position for $\phi = 20^\circ$. These results indicate that the inclination angle of the conical wire array serves as a control parameter for the propagation velocity of the jet.

The intensity of the scattered spectra decreases at higher axial positions, consistent with the electron density profile obtained. Throughout the observed range, the spectra do not exhibit ion-acoustic features, likely because $ZT_e<3T_i$\cite{froula2012plasma}, resulting in a scattered spectrum that remains nearly Gaussian. 

Due to the absence of ion-acoustic features, an exact measurement of the electron temperature ($T_e$) is not possible. Instead, the TS fitting was constrained using the electron density obtained from a combination of deflectometry and spectral data. Under these constraints, TS provides the ion temperature ($T_i$), and subsequently, an upper limit for $T_e$. Then we vary $T_e$ in the fit until the onset of a double-peaked spectral feature is observed. Based on the fitted values, a lower bound for the average ionization state ($Z$) is established

%and then we vary $T_e$ in the fit until the onset of a double-peaked spectral feature is observed to stablish an upper limit for $T_e$. .

%\textcolor{red}{Considering McWhirter criteria \cite{mcwhirter}, the plasma flow would be in local thermodynamic equilibrium (LTE) for the entire extent observed with TS. Then, Boltzmann plot was used to estimate an accessible range of electron temperatures $T_e$, thus constraining the electron temperatures to be tested in the TS spectral fit to the range $8-25 eV$. 

%In addition, the electron density measured by the combination of the deflectometry and spectral data was used to constrain the TS fitting. 

\begin{figure}[hb]
    \centering
    \includegraphics[width=0.48\textwidth]{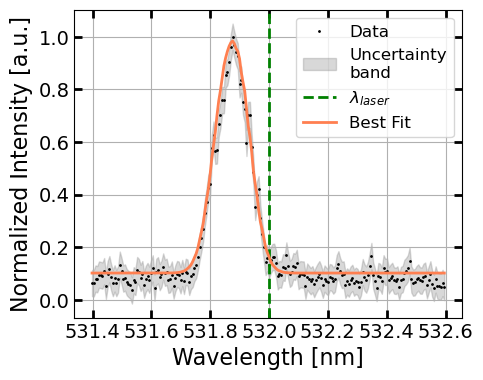}
    \caption{TS spectrum of the $40^\circ$ cone, collected by a single fiber at Z$\approx2$ mm from the anode, $\sim 450$ ns after the current discharge. The data are shown in black dots and the best fit in the continium orange line}
    \label{fit thomson}
\end{figure}

Fig. \ref{fit thomson} shows an example of the spectrum obtained and its corresponding fit for a $40^\circ$ cone, collected by a single fiber at approximately $2$ mm from the anode and $\sim 450$ ns after the current discharge.

\begin{figure}[]
    \centering
    \includegraphics[width=0.46\textwidth]{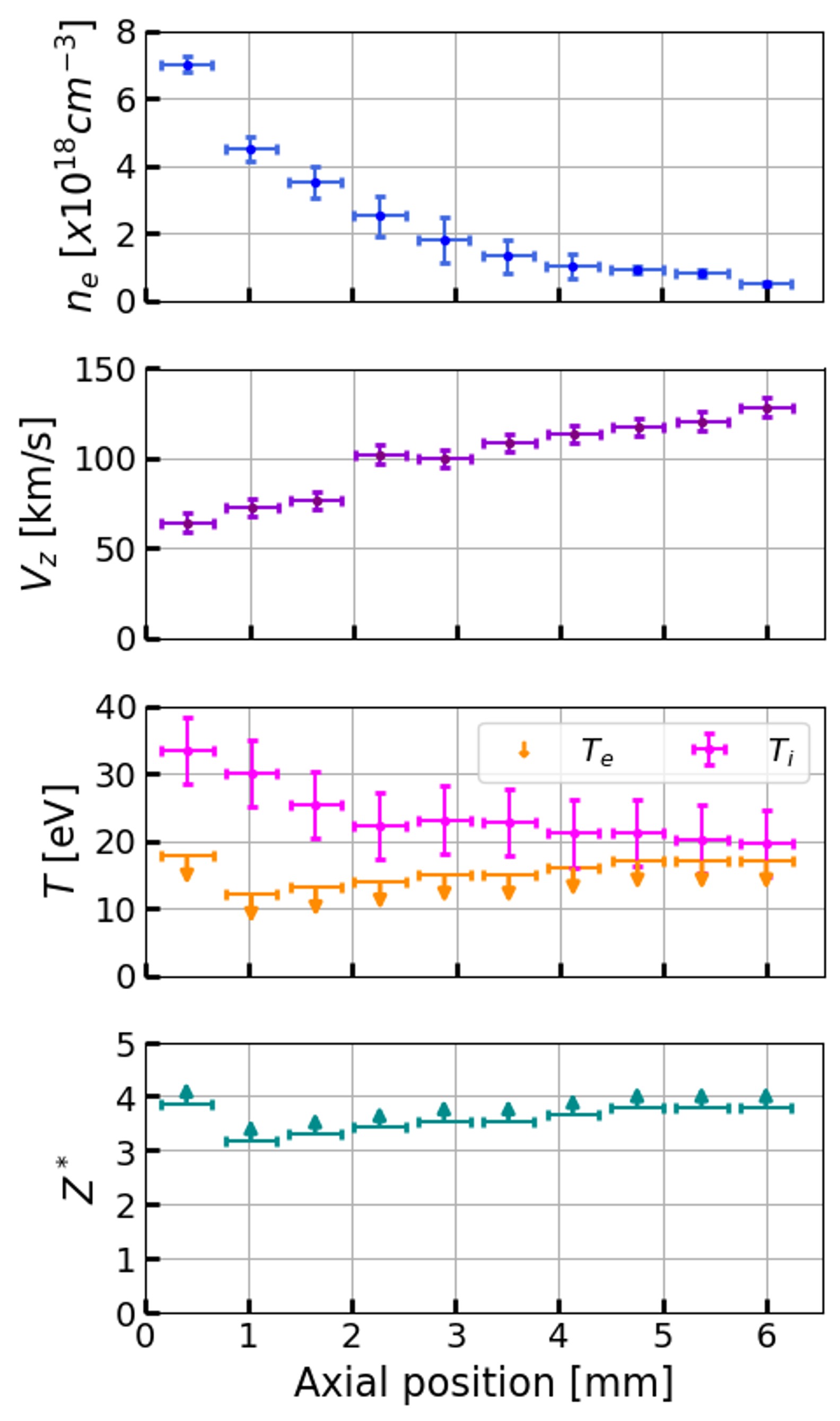}
    \caption{Axial plasma parameter profiles of the jet emitted by a 40° opening angle conical wire array, ~450ns after the beginning of current discharge.}
    \label{resultados thomson}
\end{figure}

The plasma parameters obtained for the jet emitted by a $40^o$ conical wire array are summarized in Fig.\ref{resultados thomson}.
%The product between $n_e$ and $V_z$ presented in Fig.\ref{resultados thomson} shows that $n_e$  decreases more significantly than the increasing rate of $V_z$, generating that $[\rho v]\neq 0$ along the propagation axis. %\textcolor{red}{no se conserva momento lineal? hay friccion o fuerza neta actuando}.  
It is observed that, for most axial positions, $T_i > T_e$. This temperature difference gradually decreases along the axial position until $T_e$ and $T_i$ almost converge.
The ionization state $Z$ increases slightly along the axial position, ranging from approximately 3 to 4.

\section{Discussion}

\begin{figure}[h]
    \centering
    \includegraphics[width=0.43\textwidth]{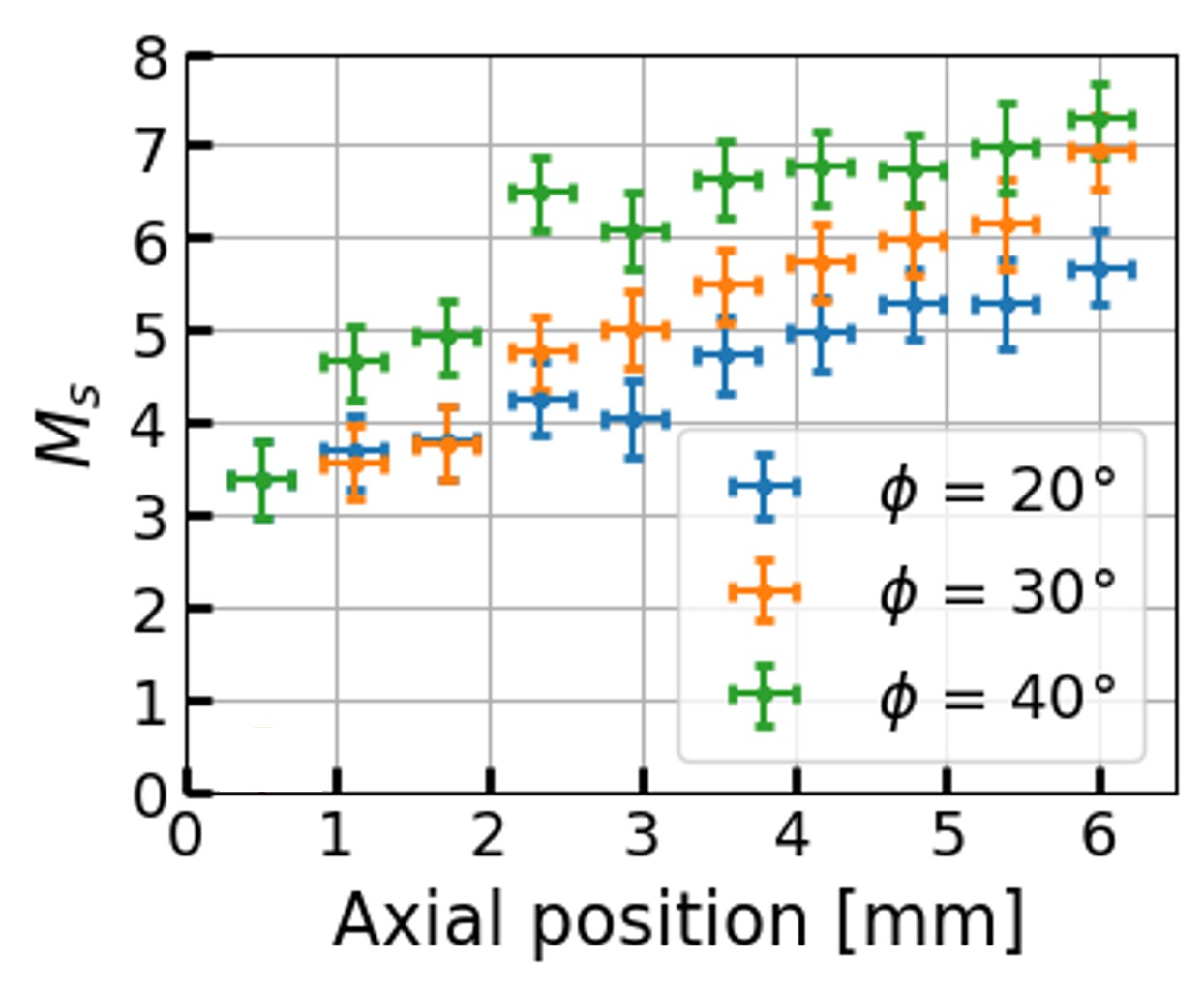}
    \caption{Axial mach numer profile determined for the plasma jet, for three different opening angles.}
    \label{Mach}
\end{figure}

The comprehensive analysis of moiré deflectometry, optical spectroscopy, and TS data allows for the evaluation of different plasma dimensionless numbers. The sonic Mach number as a function of axial position was determined for each opening angle $\phi$, as shown in Fig. \ref{Mach}. It is observed that the plasma jet remains supersonic throughout the measured axial range, and the sonic Mach number increases with $\phi$, following the relation $M_{s, \phi=20^\circ} < M_{s, \phi=30^\circ} < M_{s, \phi=40^\circ}$, in agreement with the corresponding velocity profiles.

To calculate global dimensionless numbers, we considered a set of characteristic plasma parameters of every plasma jet. These are evaluated considering $L_n=2.86$ mm as the length scale. The characteristic plasma parameters considered and the dimensionless numbers obtained are presented in table \ref{tabla parametros}.

\begin{table*}[t] % Usa table* para que ocupe ambas columnas y [b] para posicionarla abajo
    \centering
    \renewcommand{\arraystretch}{1} % Aumenta el espaciado entre filas
    \begin{tabularx}{\textwidth}{c *{4}{>{\centering\arraybackslash}X}}
 % Usa X para que las columnas se expandan
        &  &  & Value &  \\
        \textbf{Measured Parameter} & \textbf{Symbol}   & $\boldsymbol{\phi=20^\circ}$ & $\boldsymbol{\phi=30^\circ}$ & $\boldsymbol{\phi=40^\circ}$ \\ 
        \midrule
        Peak electron density [$\times 10^{18}$ $cm^{-3}$]  & $n_e$   & $1.4\pm0.6$ & $1.8\pm0.5$ & $1.6\pm0.5$  \\
        Flow Velocity [km/s] & $v$   & $71\pm5$ & $89\pm5$ & $100\pm5$   \\
        Electron temperature [eV]   & $T_e$ & $17\pm3$ & $15\pm2$ & $15\pm2$  \\
        Ion temperature [eV] & $T_i$  & $22\pm5$ & $23\pm4$ & $24\pm4$ \\
        Average ionisation & $Z^*$  & $3.7\pm0.3$ & $3.5\pm0.2$ & $3.5\pm0.2$ \\
        \\
        \textbf{Dimensionless parameter} &  &  \\ 
        \midrule
        Sonic Mach number & $M_s$ & $3.9\pm0.4$ & $5.0\pm0.5$ & $6\pm0.5$ \\
        Reynolds number [$\times 10^5$] & $R_e$ & $1.4\pm0.6$ & $1.5\pm0.6$ & $1.8\pm0.4$ \\
        Magnetic Reynolds number & $R_{e_M}$ & $22\pm4$ & $30\pm5$ & $39\pm6$ \\
        Knudsen number [$\times10^{-4}$] & $K_n$ & $3.1\pm0.3$ & $3.0\pm0.4$ & $3.4\pm0.4$ \\
        Peclet number [$\times10^3$] & $P_e$ & $5.2\pm0.3$ & $6.7\pm0.7$ & $9.0\pm0.6$ \\
    \end{tabularx}
    \caption{Characteristic plasma parameters of the different $\phi$ jets at $z\approx3$mm, $\sim 450$ ns after current start. To evaluate the Reynolds, magnetic Reynolds and Peclet number, a scale lenght given by $L_n=2.86$ mm is considered}
    \label{tabla parametros}
\end{table*}
%remains nearly constant along the jet and is notably high, approximately $6 \times 10^4$, indicating that viscous dissipation occurs at scales much smaller than the system size. A much more modest $R_{eM}$ of $\sim 10$ shows that magnetic field is expected to be advected.

%To compute the Reynolds number ($R_{e}$), Reynold magnetic number ($R_{eM}$) and Peclet number ($P_e$), a scale lenght of $6$ mm, the aproximate length of the jet, was used. Considering characteristic plasma parameters presented in \textcolor{red}{table 1}, $R_e\sim 10^4$, $R_{eM}\sim 30$, $P_{e}\sim 10^4$.

Consistent with Fig. \ref{Mach}, the plasma jet remains supersonic for all cone geometries. The large Reynolds number suggests that viscous dissipation occurs at scales much smaller than the system size, making viscous effects negligible for every $\phi$. The magnetic Reynolds number indicates that the magnetic field is primarily advected, with magnetic diffusion becoming relevant only at smaller spatial scales. In addition, it shows a slight increase with larger cone opening angles. The small Knudsen number, defined as the ratio of the ion-ion mean free path to the characteristic length scale, confirms that ions within the plasma jet are highly collisional, regardless of the cone geometry. Lastly, the Peclet number increases slightly with the opening angle but, in all cases, its high value suggests that thermal conductivity remains negligible at the characteristic spatial scale of the jet. Furthermore, relevant characteristic length scales can be determined. Table \ref{tabla escalas} presents relevant length scales calculated using the characteristic parameters of the $\phi=30^\circ$ plasma jet.

%\textcolor{brown}{The small internal ion-ion mean free path ($\lambda_{i,i}$) and ion inertial length ($d_i$) indicate that both internal collisions between ions and energy dissipation are expected to occur at scales much smaller than the characteristic length of the plasma jet ($\sim 0.1 \mu m$). In contrast, ohmic dissipation and electron heat conduction, represented by $L_\eta$ and $L_{\chi}$, take place on scales of a few tens of microns.}

To explain the observed temperature profiles, the competing mechanisms of heating, cooling, and transport are analyzed.

For an aluminium plasma in this density regime, radiative cooling becomes significant. The radiative cooling time is given by \cite{suzuki2015bow}

\begin{equation}
    \tau_{cool}=2.4\times10^{-12} \frac{(1+Z)T_e}{Zn_z\Lambda(n_i,T_e)}
    \label{t_cool eq}
\end{equation}

where $\Lambda(n_i,T_e)$ is the cooling function.

\begin{figure}[h]
    \centering
    \includegraphics[width=0.425\textwidth]{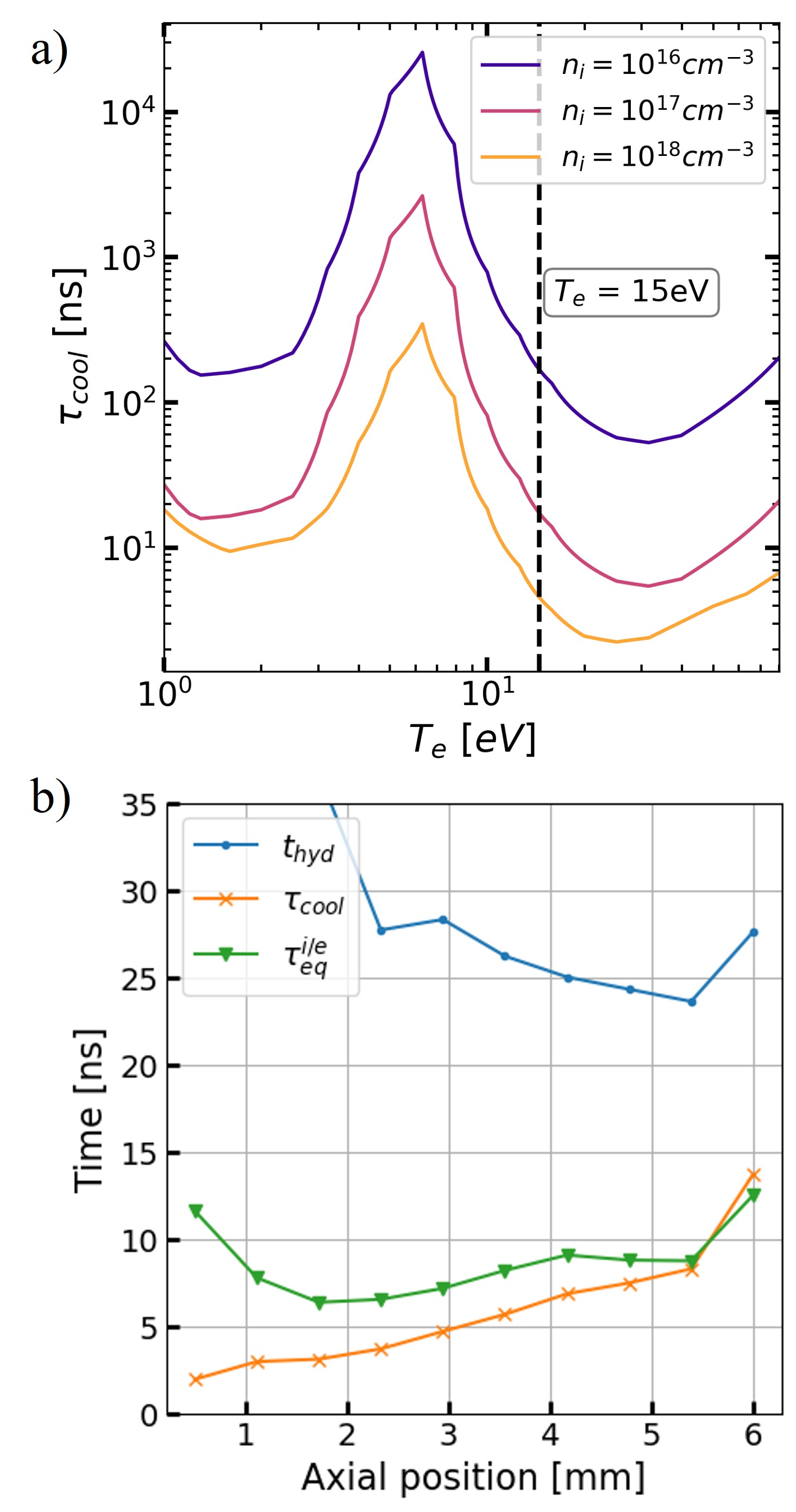}
    \caption{a) Cooling time as a function of electron temperature for aluminium at different $n_i$. b) Time scales computed with the plasma parameters determined for each axial position.}
    \label{tiempos caracteristicos}
\end{figure}

Combining numerical simulation presented in references \cite{suzuki2015bow, russell2021bow,valenzuela2024structure} with the $T_e$ and $n_i$ measured in this work, $\Lambda(n_i,T_e)$ is estimated and used to determine $\tau_{cool}$.
Fig.\ref{tiempos caracteristicos}a shows the cooling time for an aluminium plasma within the relevant electron temperature range and for different values of $n_i$.
This shows that the plasma jet is efficiently cooled across the entire electron temperature range of interest. The jet exhibits stronger cooling for higher densities, which is consistent with colder electrons at the base of the jet, where the density is larger. The cooling efficiency decreases along the axis, according with a decrease in density for higher axial positions.

%Furthermore, a characteristic electron temperature $T_e=15$eV is marked by the vertical dashed line, ilustrating that in the relevant $T_e$ range, $\tau_{cool}$ is on its way to the bottom of the curve. In this regime, an increase in the temperature of electrons enhances the radiative emission efficiency, meaning that as electrons cool, the cooling time decreases.

%Moreover, Fig.\ref{tiempos caracteristicos}b compares the cooling time with the ion-electron equilibration time at each axial position. It is observed that the relationship $\tau_{cool} < \tau_{eq}$ holds for most of the length of the jet. This indicates that electrons gain energy from ions and radiate it, albeit less efficiently at higher positions along the jet.

Moreover, Fig.\ref{tiempos caracteristicos}b compares the cooling time and ion-electron equilibration time $\tau_{eq}^{i/e}$ \cite{plasmaformulary} with the plasma hydrodynamic time ($t_{hyd} = L_n/V_z$). Since $\tau_{eq}^{i/e}<t_{hyd}$, we would expect complete thermal equilibrium between ions and electrons. However, since $\tau_{cool} < \tau_{eq}^{i/e}$, electrons cool down before thermal equilibrium with the ions is reached. This leads to thermal decoupling between electrons and ions, which is consistent with the observed temperature profiles. Therefore, the ions are indirectly cooled via energy exchange with the electrons, which radiate energy away.

By comparing the temperature difference between the electrons and the ions at each axial position, with the difference between $\tau_{cool}$ and $\tau_{eq}^{i/e}$, the temperature profile can be fully described. The maximum difference between electron and ion temperatures is reached in the most dense zone of the jet, where $\tau_{cool}<<\tau_{eq}^{i/e}$. While at higher axial positions of the jet, the thermal decoupling is minimum, where $\tau_{cool}\approx\tau_{eq}^{i/e}$. By considering this, the axial $T_e$ profile observed in the TS data agrees with variations in cooling efficiency for different densities. A decrease in electron density along the axis allows the electrons to retain their temperature, resulting in an axial increase in electron temperature.

\begin{table}[h]
    \centering
    \begin{tabular}{ccc} % 3 columnas alineadas al centro
        Scale length & Symbol & Value ($\mu$m) \\ % Encabezadosaparte 
        \midrule
        Internal ion-ion mean free path & $\lambda_{i,i}$ & $0.13$ \\
        Resistive diffusion length  & $L_{\eta}$  & $270$ \\
        Electron thermal diffusion length & $L_{\chi}$  & $90$ \\
        Ion inertial length  & $d_i$ & $0.3$ \\
  
    \end{tabular}
    \caption{Characteristic scale lengths calculated with the characteristic parameters of the $\phi=30^\circ$ plasma jet presented in table \ref{tabla parametros}}
    \label{tabla escalas}
\end{table}

\section{Conclusions}

In this work, we have performed a comprehensive experimental study of plasma jets emitted by aluminium conical wire arrays driven on the Llampudken pulsed power generator. By combining three complementary diagnostics, including moiré schlieren deflectometry, optical emission spectroscopy, and Thomson scattering we achieved spatially resolved measurements of key plasma parameters, including electron density, flow velocity, and ion/electron temperatures.\\

Our results indicate that the inclusion of a physical aperture at the top of the conical array played a crucial role in isolating the jet from ablated plasma streams coming from the wires themselves. This modification altered the overall density profile observed in previous experiments using intererometry, revealing a jet with significant lower density than previously reported. We consider that these previous measurements have overestimated electron densities by adding both jet and ablation stream densities as a single flow. 

For the jet itself, we have experimentally shown a exponential decay in axial electron density distribution
%One of the key findings is the measurement 
with a characteristic axial scale length, \(L_n\), which is consistent across different opening angles. Besides that, we found that as density and temperature profiles remain largely invariant with the array opening angle, the jet propagation velocity increases with \(\phi\), providing a straightforward way to control this parameter. The plasma jets are found to be supersonic, highly collisional, and in a regime where magnetic and thermal diffusion are negligible at the characteristic spatial scales. The axial temperature profile is shaped by the balance between ion-electron energy exchange and radiative cooling, with thermal decoupling appearing near the jet base due to higher density and enhanced radiative losses.

Since pulsed power plasma jets have often been used as laboratory astrophysics platforms, a more detailed description of internal parameters and their gradients along the plasma jet support their relevance in scalability to astrophysical systems such as young stellar object outflows and offer valuable benchmarks for simulations. As collisionality in plasmas is highly dependant on the velocity, these results also encourage the use of conical wire array jets as platform to study plasma collisionality taking advantage of known velocity values, their gradients, and the $\phi$ parameter as a way to control them.

Future work will also aim to expand the diagnostic capabilities by incorporating magnetic field measurements —specifically Zeeman spectroscopy— which will enable the determination of magnetic pressure, thermal and ram beta, Alfvénic Mach number, and additional length scales such as the ion and electron gyroradii. These measurements will further strengthen the connection between laboratory plasma jets and astrophysical outflows, allowing for more complete and accurate comparisons with theoretical models and simulations, as well as enabling the study of plasma–plasma interactions and the characterization of their collisionality regimes.

\section{Acknowledgements}
This work has been partially funded by Fondecyt/Regular 1231286, 1220533, 3230401 and 1211131 projects. L. Izquierdo acknowledges doctoral funding from ANID-Subdirección de Capital Humano/Doctorado Nacional/2023-21230431.

\bibliographystyle{ieeetr}         % o cualquier otro estilo válido
\bibliography{references}         % este es tu archivo .bib, sin extensión

\end{document}